\documentstyle[11pt]{article}

\advance\voffset by -2.2cm
\advance\hoffset by -2cm
\def\beq{\begin{equation}}
\def\eeq{\end{equation}}
\def\bqa{\begin{eqnarray}}
\def\eqa{\end{eqnarray}}

\newtheorem{conj}{Conjecture}
\def\({\left(}
\def\){\right)}
\def\Ref#1{(\ref{#1})}

\def\Sig{\Sigma}
\def\tr{{\rm{tr}}}

\def\frac#1#2{{#1 \over #2}}

\def\pdd#1#2{{\partial #1\over\partial #2}}
\def\dd#1#2{{d #1\over d #2}}
\def\pD#1d#2#3{{\partial^{#1} #2 \over \partial #3^{#1}}}
\def\DD#1d#2#3{{d^{#1} #2 \over d #3^{#1}}}

\def\a{\alpha}
\def\b{\beta}

\def\d{\delta}
\def\e{\eta}
\def\l{\lambda}
\def\m{\mu}
\def\n{\nu}

\def\z{\zeta}
\def\A{{\cal{A}}}  %
\def\B{{\cal{B}}}  %

\def\Mt{{\tilde{M}}}  %
\def\Perm{{\cal P}}

\begin{document}

\hfill DAMTP 94-17
\vskip 24pt
\centerline{\bf Classical Functional Bethe Ansatz for $SL(N)$: }
\vskip 8pt
\centerline{\bf  Separation of Variables for the Magnetic Chain}
\vskip 24pt
\centerline{D.R.D. Scott\footnote{e-mail:D.R.D.Scott@amtp.cam.ac.uk}}
\vskip 12pt
\centerline{\it Department of Applied Mathematics and Theoretical Physics}
\centerline{\it University of Cambridge}
\centerline{\it Silver Street, Cambridge CB3 9EW, U.K.}
\vskip 50pt
\centerline{\bf Abstract}
\vskip 24pt
 The Functional Bethe Ansatz (FBA) proposed by Sklyanin is a
method which gives separation variables for systems for which an
$R$-matrix is known. Previously the FBA was only  known for $SL(2)$ and $SL(3)$
(and associated) $R$-matrices. In this paper I advance Sklyanin's
program by giving the FBA for certain systems with $SL(N)$
$R$-matrices. This is achieved by constructing rational functions
$\A(u)$ and $\B(u)$ of the matrix elements of $T(u)$, so that, in the
generic case, the zeros $x_i$ of $\B(u)$ are the separation
coordinates and the $P_i=\A(x_i)$ provide their conjugate momenta. The
method is illustrated with the magnetic chain and the Gaudin model,
and its wider applicability is discussed.
\vfill
\vskip -12pt
March 1994 \hfill
\pagestyle{empty}
\eject
\pagenumbering{arabic}
\pagestyle{plain}

\baselineskip 16pt plus 3pt minus 3pt

\section{Introduction}

\subsection{Separation of variables in Integrable Systems}

A classical system with a  $2n$-dimensional phase space is said to be
{\sl integrable} if there exist $n$ independent functions on the phase
space (called integrals of motion) which Poisson commute among
themselves and with the Hamiltonian.

Classically a system is said to be {\sl separable} if there exist variables
in which the Hamilton-Jacobi equation separates. Traditionally this
means introducing $n$ separation constants so that the Hamilton-Jacobi
equation (involving all $n$ pairs of conjugate variables) can be
replaced by $n$ equations involving one each (and possibly a further equation
constraining the allowed separation constants at each energy).

Liouville's Theorem~\cite{viA:book} says that for integrable systems there
exist
action-angle variables (at least locally) in which the Hamiltonian is
a function of the action variables alone and hence separates. Thus for
integrable systems it is known that separation variables exist (at
least locally). However the formula for constructing action-angle
variables (which involves integrating over invariant tori) is not
tractable in general.

What is needed is an explicit method for constructing
separation variables for arbitrary integrable systems. Two such methods exist.
One uses algebraic geometry and has been applied to  loop algebras with
linear Poisson brackets~\cite{AHH:DC}. The other, the Functional
Bethe Ansatz (FBA) works for systems with $SL(2)$ or $SL(3)$ $R$-matrices of
certain forms. In this paper I extend the FBA to $SL(N)$ for $N>3$. I
work with the magnetic chain and Gaudin magnet, but the method and
results are much more widely applicable as I will discuss at the end
of the paper. Although this extension has equations in common
with the algebraic geometric technique no knowledge of algebraic
geometry is required to understand this paper.

 Some insight may be gained from considering the reverse problem (i.e.
how to construct integrals of motion given a separation) which is much better
understood. Jacobi's Theorem~\cite{viA:book} can be interpreted as
saying that if a system separates then it is integrable, with the
integrals of motion corresponding to the separation constants. Thus
given a separation of a Hamiltonian there is a systematic method for
constructing the integrals of motion, and the separation of the
original Hamiltonian induces a separation of these integrals of
motion. Thus it is natural in integrable systems to think of
separating the system without any reference to a particular integral
of motion. Separation of variables may then be (re)defined as seeking
variables $x_i$ and $p_i$, with canonical Poisson brackets, which
satisfy $n$ separated equations,
\beq
\Phi_j(x_j,p_j,I_1,\ldots,I_n)=0 \quad j=1,\ldots,n
\eeq
where in these equations the dependence on separation constants has
been replaced with dependence on the integrals (which are of course
{\sl constants} of the motion).
This is the starting point used by Sklyanin~\cite{ekS:SL3MC} for his
Functional Bethe Ansatz.

\subsection{$R$-matrices and Lax Pairs}

The two methods already proposed and this paper all use the
matrix formalism associated with $R$-matrices and Lax Pairs. To
understand this paper only the basics of this formalism need be known,
further details may be found for example in
\cite{amP:book,jAmT:GRmIS,maSTS:CRm}.

In the $R$-matrix approach the operator content of the theory is
contained in an $N\times N$ matrix
function on the phase space $T(u)$ depending on a spectral parameter
$u$, the algebraic structure of the theory is given by the so called $R$-matrix
algebra. This can be quadratic,
\beq \{ \stackrel{1}{T}(u),\stackrel{2}{T}(v) \} =  [
R(u-v),\stackrel{1}{T}(u)\stackrel{2}{T}(v) ] \eeq
or linear,
\beq\{ \stackrel{1}{T}(u),\stackrel{1}{T}(v) \} =[ R(u-v),
\stackrel{1}{T}(u) + \stackrel{2}{T}(v) ] \quad.\label{eq:LRm}\eeq
Here $\stackrel{1}{T}$ denotes $T(u)\otimes 1$ etc.\ and $R$ is an
$N^2\times N^2$ matrix acting in the tensor product.

The $R$-matrix must obey a consistency condition, the Classical
Yang-Baxter equation.
 In the case of Lax Pairs we have a pair of matrices $L(u)$ and
$M(u)$ depending on spectral parameter $u$, $L(u)$ is a function on
the phase space, while $M(u)$ is usually a constant.
$L$ and $M$ give  a Lax representation of a system if the equations of
motion of the system are equivalent to the following evolution
equation for $L$,
\beq \dd{L(u)}{t} = [ L(u),M(u) ] \quad .\eeq
{}From this equation it is clear that the quantities generated by the
spectral invariants of $L(u)$ are constants of the motion, but a
priori they do not Poisson commute. The necessary and sufficient
condition that the eigenvalues Poisson commute is that the Lax matrix
obeys an $R$-matrix type Poisson brackets \cite{oBcmV:HSLe}, which for
antisymmetric $R$-matrices reduces to equation~\Ref{eq:LRm}.

 This paper uses the $R$-matrix
formalism. What is important to know for this paper is that in general
the spectral invariants of $T(u)$ provide the integrals of motion. So
long as they give rise to enough the system is integrable.

\subsection{The Method}            \label{ss:key}
The search for separation variables can be thought of as happening in
two stages, (1) look for variables which give rise to separated
equations, then (2) check that these variables have the correct
commutation relations.

Crucial to both methods is the idea that if
$\z_i$ is an eigenvalue of $T(u_i)$ then $u_i$ and $\z_i$
automatically obey the following `separated' equation,
 \beq \det\left( \z_i - T(u_i) \right) =0 \quad . \label{eq:ev}\eeq
This is indeed a separated equation in that it only depends on $u_i$,
$\z_i$ and the spectral invariants of $T(u)$ (i.e. the integrals of
motion). Any such $u_i$ and $\z_i$ found in this way therefore
complete step (1). The problem is to choose them in such a way that
they have the correct Poisson brackets and form a basis of the phase space.

\subsubsection{Functional Bethe Ansatz}

The Functional Bethe Ansatz was proposed by Sklyanin
\cite{ekS:SL3MC,ekS:GM,ekS:FBA,ekS:QISM} as a
method of finding separation variables for integrable systems, this
paper is only concerned with the classical version. It is a
blending of ideas from the Bethe Ansatz and separation of variables.
It is applicable to systems for which an $R$-matrix (or indeed Lax)
representation of a certain form is known.

The FBA was first proposed for the case of $2\times 2$ matrices.
In this case the basic objects are
\beq T(u) = \pmatrix{ A(u) & B(u) \cr C(u) & D(u) } ,\eeq
and these obey $R$-matrix relations. The Functional Bethe Ansatz says
that if the  $\{ x_i, i=1,\ldots,M\}$ are the zeros of $B(u)$ then the $x_i$
provided a set of separation coordinates and the set of $P_i$ given by
$P_i=A(x_i)$ are their conjugate momenta (or possibly a function of
them). The separated equations are then,
\beq \det\left( P_i - T(x_i) \right) =0 \quad  . \eeq
It has been successfully applied in this ($2\times 2$) case to a wide
class of systems. The FBA for $SL(3)$ was studied by Sklyanin
in~\cite{ekS:SL3MC} for the $SL(3)$-magnetic chain and the Gaudin model.
In this paper he conjectures how the method might be extended to
$SL(N)$. For the magnetic chain the conjecture is as follows (with a
similar one for the Gaudin magnet but with canonical Poisson brackets),
\begin{conj}                
There exist functions $\A$ and $\B$ on $GL(N)$ such that
the following two assertions are true. Firstly that $\A(T)$ is an algebraic
function and $\B(T)$ is a polynomial of degree
$D=MN(N-1)/2$ in the matrix elements $T_{mn}$. Secondly that the variables
$x_j, P_j \quad (j=1,\ldots,D)$ defined from the equations,
\beq \B(T(x_i))=0, \quad\quad P_j=\A(T(x_j)) \eeq
have Poisson brackets
\beq \{x_j,x_k\}=\{P_j,P_k\}=0, \quad\quad \{P_j,x_k\}=P_j\d_{jk} \eeq
and, besides, are bound to the Hamiltonians (integrals of motion) by
the relations
 \beq \det\left( P_i - T(x_i) \right) =0 \quad . \eeq
\end{conj}                  
The scheme proposed by Sklyanin to obtain $\A$ and $\B$ is to put $T$ into
block triangular form  by using a similarity transform
\beq T'=K^{-1}TK \eeq
 where $K$ is a matrix depending on some parameters $k_1,\ldots k_Q$.
(I shall show that $Q$ need only be $N-2$.)
The $k_i$ should then be eliminated from the resulting equations to
leave a single equation $\B(u)=0$. No similarity transformation is
required for $N=2$ and the similarity transformation for $N=3$ was
performed by Sklyanin in Ref\cite{ekS:SL3MC}, where he proves the
commutation relations for the magnetic chain and Gaudin magnet.

\subsubsection{Algebraic Geometric Method}

M.R. Adams, J.Harnard and J.Hurtubise \cite{AHH:DC} have used algebraic
geometric methods to find separation coordinates for systems that have
Lax representations with linear $R$-matrices and Lax matrices of the
form,
\beq L(u) = uY + u\sum_{i=1}^{M} \frac{L_i}{u-\a_i} \eeq
with $Y\in gl(n)$ and $\a_i$ as complex constants (it also works
for multiple poles).

 In this approach separation
variables are constructed as the generically distinct finite solutions of
the equation,
\beq \tilde{M} (u,\z)V_0 =0 \label{eq:ahh}\eeq
where  $\tilde{M}$ is the classical adjoint of
\beq M(u,\z) := T(u) - \z I \eeq
The solutions of this equation give Darboux coordinates
in systems with linear Poisson brackets of a certain from. The
solutions of this equation are again bound to the integrals of motion
by equation~\Ref{eq:ev}. As a defining equation for the separation
variables equation~\Ref{eq:ahh} has the advantage that (because it
can be treated in the language of algebraic geometry) degenerate cases
can be handled easily. Perhaps, most importantly, the commutation
relations can be straightforwardly calculated.

A disadvantage as compared with Sklyanin's scheme is that the
defining equation involves two parameters (the spectral
parameter and an eigenvalue parameter).

\subsection{Overview of Paper}

 In this paper I advance Sklyanin's program by
giving the FBA for $SL(N)$. This is done by constructing polynomials $\A$ and
$\B$ in the matrix elements, so that generically the zeros $x_i$ of $\B(u)$
give the separation coordinates and the $P_i=\A(x_i)$ provide their conjugate
momenta in the cases of the magnetic chain and Gaudin model. This
paper deals primarily with the magnetic chain which is reviewed in
section~\Ref{sec:MC} (the Gaudin model being dealt with at the end
(section~\Ref{sec:GM}).)

In section \Ref{sec:AB} I obtain candidates for $\A$ and $\B$ for the case
of $N\times N$ matrices, in the sense that produce variables that give
rise to separated equations, without however any consideration of
whether they have the correct commutation relations at this stage.
Sklyanin's program of similarity transforms is used.

In section \Ref{sec:CR} it is shown that {\it generically} $\A$ and
$\B$ give the same separation coordinates as obtained from
equation~\Ref{eq:ahh}, and then this equivalence is used to prove the
commutation relations for the magnetic chain. In doing this the
crucial calculation in Ref.~2 is extended to show that
equation~\Ref{eq:ahh} gives separation variables in the case of quadratic
Poisson brackets  (given by the permutation $R$-matrix,
$\frac{\Perm}{u-v}$ see equation~\Ref{eq:PRm}).

In section \Ref{sec:GM} it is shown how the results carry over to
the (linear $R$-matrix) case of the Gaudin Model, and to systems with
more general $R$-matrices.

\section{The Magnetic Chain} \label{sec:MC}

The variables of the  non-homogeneous classical $SL(N)$ magnetic
chain are $S_{\a\b}^{(m)}
,\a,\b=1,\ldots,N;m=1,\ldots,M$ (where $M$ is the length of the spin
chain). The variables are not completely independent but a related by
$\sum_{\a=1}^{N} S_{\a\a}^{(m)}=0$.
They obey the following Poisson brackets
\beq \{ S_{\a_1\b_1}^{(m)}, S_{\a_2\b_2}^{(n)} \} = \(
S_{\a_1\b_2}^{(m)}\d_{\a_2\b_1} -  S_{\a_2\b_1}^{(m)}\d_{\a_1\b_2} \)
\d_{mn} \label{eq:spb} \eeq
which define the Kirillov-Kostant Poissonian structure on the direct
product of $M$ coadjoint orbits of $SL(N)$. The center of the algebra
is generated by the eigenvalues $\l^m_\a$ of the matrices $S^{(m)}$
\beq
\det \( S^{(m)}-\l \) = \prod_{\a=1}^{N} \(\l^m_\a-\l\). \quad
\sum_{\a=1}^{N} \l^m_\a =0  \label{eq:man}
\eeq
I shall fix the orbit, by taking $\l^m_\a$ to be fixed numbers.
Furthermore I shall assume that I have a generic orbit by requiring
the
eigenvalues of $S^{(m)}$ are distinct. The manifold defined by
equations~\Ref{eq:man} and having dimension $MN(N-1)$, is then equipped with
a non-degenerate Poisson bracket~\Ref{eq:spb}.

The monodromy matrix may be defined as,
\beq T(u)=Z(u-\d_M+S^{(M)})\cdots(u-\d_2+S^{(2)})(u-\d_1+S^{(1)}) \eeq
where $Z$ is an $N\times N$ number matrix with distinct eigenvalues
and $\d_m$ are some fixed numbers, and $u$ is the spectral parameter.

The matrix elements of $T$ are polynomial in $u$ of degree $M$ (length of the
magnetic chain). The $T(u)$ obey the following quadratic R-matrix relations,
\beq \{ T_{\a_1\b_1}(u),T_{\a_2\b_2}(v)\} = \frac{1}{u-v} \big(
T_{\a_2\b_1}(u) T_{\a_1\b_2}(v) - T_{\a_1\b_2}(u)T_{\a_2\b_1}(v) \big)
.\eeq
or in the formalism used earlier,
\bqa \{ \stackrel{1}{T}(u),\stackrel{2}{T}(v) \} &=&
[R(u-v),\stackrel{1}{T}(u)\stackrel{2}{T}(v) ] \cr
R(u) &=& \frac{\Perm}{u} \eqa
where $\Perm$ is the permutation matrix in the tensor product i.e.,
\beq R(u) = \frac{1}{u} \sum_{i,j=1}^N e_{ij} \otimes e_{ji} \label{eq:PRm}\eeq
where the $e_{ij}$ form the usual basis for $N\times N$ matrices with
a $1$ in the $ij$th position and zeros elsewhere.
The spectral invariants $t_\nu(u)$ of the matrix $T(u)$ may be defined
as the elementary symmetric polynomials of its eigenvalues,
$t_\nu(u)\equiv \tr \wedge^\nu T(u), \nu=1,\ldots,N$.
$t_N(u)=\det (T(u))$ contains the central elements and is therefore
taken to be a constant function of $u$. The non-leading coefficients of the
remaining invariants provide a commuting family of $MN(N-1)/2$
independent Hamiltonians (see e.g. \cite{ekS:SL3MC}), since this is half the
dimension of the phase space the system is integrable.

\section{$\A,\B$ in the $SL(N)$ case}  \label{sec:AB}

In this section I use the similarity transformation equations
to show that the following are candidates for $\A$ and $\B$, i.e.
that the $x_i$ and $P_i$ derived from these equations allow separation
(the proof that these variables obey the correct commutation relations
is postponed to the next section.)
\bqa \A(T(u)) &=& \epsilon_{i_1 i_2 \cdots i_{N-1}} T_{i_1N}
T^2_{i_2N} \cdots  T^{N-2}_{i_{N-2}N}
\frac{T_{i_{N-1}1}}{\det(M)} \\
\B(T(u)) &=&  \epsilon_{i_1 i_2 \cdots i_{N-1}}  T_{i_1N}
T^2_{i_2N} \cdots  T^{N-2}_{i_{N-2}N}
T^{(N-1)}_{i_{N-1}N}
\eqa
In the case the case of the magnetic chain, where $T(u)$ is a
polynomial in $u$ of degree $M$, $\B(u)$ is generically degree
$MN(N-1)/2$ insuring that it defines the correct number of separation
variables.

In this section indices labeled $i,j,k$ range from $1$ to $N-1$,
indices labelled $m,n$ range from $2$ to $N-2$ and repeated indices
are summed over their appropriate ranges.

Let
\beq K=\pmatrix{ 1 & k_2 & k_3 & \cdots & k_{N-1} & 0 \cr
                 0 &  1  & 0  & \cdots & \cdots  & 0  \cr
                \vdots  & \ddots & \ddots  & \ddots &         & \vdots  \cr
               \vdots   &     & \ddots & \ddots & \ddots & \vdots  \cr
               \vdots   &     &     & \ddots      & \ddots  & 0  \cr
                0 &  \cdots & \cdots & \cdots &   0      & 1 }
.\eeq

The requirement that,
\beq KTK^{-1} =  \pmatrix{ \A(u) & 0 & \cdots & 0 \cr
                             \#  & \cdots & \cdots  &\#  \cr
                    \vdots &  &          & \vdots  \cr
                  \# & \cdots & \cdots & \# } \eeq
where $\#$ denotes arbitrary matrix elements, implies that,
\beq \A(T(u))=  k_i T_{i1} \eeq
and that the $k_i$'s must satisfy the following set of $N-1$ equations,
\bqa
k_iT_{im} - k_m k_i T_{i1} &=& 0\quad (m=2,\ldots,N-1) \label{qce} \\
k_iT_{iN} &=& 0 \eqa
where $i$ is summed from $1$ to $N-1$.
These are $N-1$ equations in $N-2$ unknowns our aim is to eliminate
the unknowns and obtain a single consistency equation $\B(T(u))$=0.
The first $N-2$ equations are quadratic and only the last is linear,
however an equivalent set of linear equations can be obtained from
these.  Multiplying the equations \Ref{qce} by $T_{mN}$ and summing
from $2$ to $N-1$, one obtains,
\beq k_jT_{jm}T_{mN} - k_mT_{mN} k_i T_{i1} = 0 \eeq
and the last linear equation may be used to replace the last term by
$k_i T_{i1} T_{1N}$, thus
\beq k_j T_{jk} T_{kN} =0 . \label{l2e} \eeq
Now we have two linear equations, further equations can now be
obtained by iterating this procedure (e.g. to obtain the next equation
in the series multiply each of the original quadratic equations by
$T_{mi} T_{iN}$ and sum them using \Ref{l2e} to eliminate the
quadratic terms).
Introducing the notation
\beq \Sig^{(a)}_{ik} = T_{ij} \Sig^{(a-1)}_{jk},\quad
\Sig^{(1)}_{ik}= T_{ik}   \eeq
 the linear equations read,
\beq k_i \Sig^{(a)}_{iN} = 0 \label{hle} .\eeq

The first $N-2$ of these equations are sufficient to determine the
$k_i$'s, and may be written in matrix form as,
\beq M_{mn}k_n = - \Sig^{(m)}_{1N} \eeq
where $M$ is the $(N-2)\times(N-2)$ matrix,
\beq M_{mn}= \Sig^{(m-1)}_{nN} .\eeq
Hence,
\beq \det(M) k_m =  N_{mn} \Sig^{(n)}_{1N} \eeq
where $N$ is the transpose of the matrix of cofactors.
Multiplying the L.H.S. of equation\Ref{hle} (with $a=N-1$) by
$\det(M)$ and substituting,
\bqa \det(M) k_i \Sig^{(a)}_{iN} &=& \det(M)\Sig^{(N-1)}_{1N} + N_{mn}
\Sig^{(m)}_{1N} \Sig^{(N-1)}_{nN} \\
&=& \epsilon_{i_1i_2\cdots i_{N-1}} \Sig^{(1)}_{i_1N} \Sig^{(2)}_{i_2N}
\cdots  \Sig^{(N-1)}_{i_{N-1}N} \eqa
this is chosen to be $\B$.
Likewise $\A$ is obtained by eliminating the $k_i$'s, giving,
\bqa \A(T(u)) &=& \epsilon_{i_1i_2\cdots i_{N-1}} \Sig^{(1)}_{i_1N}
\Sig^{(2)}_{i_2N} \cdots  \Sig^{(N-2)}_{i_{N-2}N}
\frac{T_{i_{N-1}1}}{\det(M)} \\ \B(T(u)) &=&  \epsilon_{i_1i_2\cdots
i_{N-1}} \Sig^{(1)}_{i_1N}
\Sig^{(2)}_{i_2N} \cdots  \Sig^{(N-2)}_{i_{N-2}N}
\Sig^{(N-1)}_{i_{N-1}N} \eqa

In these expressions each index is summed from $1$ to $N-1$ however
$\A$ and $\B$ remain unchanged if the sums are extends to sums from
$1$ to $N$, thus
\bqa \A(T(u)) &=& \epsilon_{i_1 i_2 \cdots i_{N-1}} T_{i_1N}
T^2_{i_2N} \cdots  T^{N-2}_{i_{N-2}N}
\frac{T_{i_{N-1}1}}{\det(M)} \\
\B(T(u)) &=&  \epsilon_{i_1 i_2 \cdots i_{N-1}}  T_{i_1N}
T^2_{i_2N} \cdots  T^{N-2}_{i_{N-2}N}
T^{(N-1)}_{i_{N-1}N}
\eqa
To see this notice that
\beq
\Sig^{(r)}_{iN}= (T^r)_{iN} + (T^{r-1})_{iN} F^{(1)} + \cdots + T_{iN}
F^{(r-1)}\eeq and that in $\A$ and $\B$ these appear within
antisymmetric sums so only the first terms contribute.

\section{Proof of commutation relations}   \label{sec:CR}

Summation convention is NOT used in this section.

In this section I prove that the $x_i$ and $P_i$ given by $\B(x_i)=0$
and $P_i=\A(x_i)$ have the following commutation relations,
\bqa \{x_i,x_j \} = \{P_i, P_j\}=0 \\
     \{x_i,P_j\} = P_i\d_{ij} \quad .\eqa
The proof works in two stages first I show that the $(x_i,P_i)$ can be
equivalently defined as the solutions of another equation. Then this
equation is used to calculate the Poisson brackets.

\subsection{Equivalent defining equation}

In this subsection I show that (if the $\B$ has the correct number of
zeros then) then ${(x_i,P_i)}$ defined by $\A$ and $\B$ can be
equivalently defined as the generically distinct finite solutions of
the equation,
\beq \tilde{M} (u,\z)V_0 =0 \label{eq:ahh2}\eeq
where  $\tilde{M}$ is the classical adjoint of
\beq M(u,\z) := T(u) - \z I \eeq
It is very natural to think of the separation variables as being
defined by these equations since they do give eigenvalues and moreover define
the variables in terms of things for which we can calculate Poisson
brackets.  The solutions of this equation were shown to give Darboux
coordinates in systems with linear Poisson brackets \cite{AHH:DC}.

To see this consider,

\beq M'_i(\z)=K(u_i)M(u_i,\z)K^{-1}(u_i) =  \pmatrix{ P_i-\z & 0 & 0
& \cdots & 0  & 0 \cr
                       \#  & \cdots  & \cdots   & \cdots & \cdots  &\#  \cr
                    \vdots &  &   &        &         & \vdots  \cr
                  \# & \cdots & \cdots & \cdots & \cdots & \# } \eeq

where $\#$ denotes arbitrary matrix elements. Factors of $(P_i-\z)$
may be pulled out of all but one of the non-zero cofactors,
\beq
\tilde{M'}_i(\z) =  \pmatrix{  \# & (P_i-\z)c_i  \cr
                               0  & (P_i-\z)B_{ij}  \cr  } \eeq
where $c_i$ and $B_{ij}$ are the relevant (determinant) factors that multiply
$(P_i-\z)$ in each cofactor.

Thus for generic $V_0$ (i.e. one for which $V_0^1=0$), $z_i=P_i$
solves the equation,
\beq \tilde{M'}(u_i,z_i) V_0 = 0 \eeq
and hence $\l=u_i$ and $\z=P_i$ are generic solutions of the
equivalent equation
\beq \tilde{M}(\l,\z) V_0 = 0 \eeq

In the case that $\B$ does not give as many solutions as \Ref{eq:ahh2}
this proof breaks down. Indeed if $\B$ does not give the same number
of zeros as half the dimension of the phase space then the method can
not be applied in its present form. For example the $R$-matrices of
Kuznetsov~\cite{vbK:QR} (for the reducible systems of Kalnins et al.
\cite{egK:book,BKM:sshl}) cannot
be used to define separation variables in this manner. It would be
very satisfying to have a FBA type method that could handle such
degenerate $R$-matrices.


\subsection{Proof of commutation relations}

Since the $x_i,P_i$ are given as the generic solutions of
equation~\Ref{eq:ahh2} we may choose $(V_0)_i= \d_{iN_0}$ where $N_0$ is
fixed but arbitrary (and could be chosen to be $2,3,\ldots,N-1$ or
$N$). Then $(x_i,P_i)$ are given by the conditions,
\beq \Mt_{kN_0}(\l,\z) = 0 \eeq
Generically these points are specified by just two of these, choose
\beq \Mt_{1N_0}=\Mt_{2N_0}=0 \eeq
and the matrix
\beq F_\nu := \pmatrix{ \pdd{\Mt_{1N_0}}{u} & \pdd{\Mt_{1N_0}}{\z} \cr
 \pdd{\Mt_{2N_0}}{u} & \pdd{\Mt_{2N_0}}{\z} } (u_\nu, \z_\nu) \eeq
is invertible. Hence
\bqa \pmatrix{ \{u_\n,u_\m\} & \!\!\!\{u_\n,\z_\m \} \cr
               \{\z_\n,u_\m\} & \!\!\!\{\z_\n,\z_\m \} } &=&  \nonumber\\
  && \hskip -75pt
{F_\n}^{-1}
\pmatrix{ \{ \Mt_{1N_0}(u_\n,\z_\n), \Mt_{1N_0}(u_\m,\z_\m) \} &\!\!\!
          \{ \Mt_{1N_0}(u_\n,\z_\n), \Mt_{2N_0}(u_\m,\z_\m) \} \cr
          \{ \Mt_{2N_0}(u_\n,\z_\n), \Mt_{1N_0}(u_\m,\z_\m) \} &\!\!\!
          \{ \Mt_{2N_0}(u_\n,\z_\n), \Mt_{2N_0}(u_\m,\z_\m) \}      }
(F_\m^T)^{-1} \label{eq:pbm}\eqa
The Poisson brackets of the adjoint can be calculated from the Poisson
brackets of $M$ by using the derivation property,
\beq
\{ \Mt_{ij}(u,\z),\Mt_{kl}(v,\e) \} = \sum_{pqrs}
\pdd{\Mt_{ij}(u,\z)}{M_{pq}(u,\z)} \pdd{\Mt_{kl}(v,\e)}{M_{rs}(v,\e)}
\{ M_{pq}(u,\z),M_{rs}(v,\e)\} \label{eq:pbd} \eeq
Now
\beq \{M_{pq}(u,\z),M_{rs}(v,\e) \} = \frac{1}{u-v}\(
T_{rq}(u)T_{ps}(v)- T_{ps}(u)T_{rq}(v) \) \eeq
and it follows from $ \Mt(u,\z) M(u,\z) = \det \(M(u,\z)\) I $ that
\beq \pdd{\Mt_{ij}(u,\z)}{M_{pq}(u,\z)} = { { \Mt_{qp}(u,\z)
\Mt_{ij}(u,\z) - \Mt_{ip}(u,\z)\Mt_{qj}(u,\z) }\over{\det M(u,\z) }}
\label{eq:pdc}\eeq
Substituting these into \Ref{eq:pbd} one obtains,
\bqa
\{ \Mt_{iN_0}&,&\Mt_{kN_0} \} = \sum_{pqrs}
\( \d_{qs}\Mt_{iN_0}(u)-\d_{is}\Mt_{qN_0}(u)
\)\(\d_{sq}\Mt_{kN_0}(v)-\d_{kq}\Mt_{sN_0}(v)\)  \nonumber\\
&-& \( \d_{rp}\Mt_{iN_0}(u)-\d_{rN_0}\Mt_{ip}(u)\)
\(\d_{pr}\Mt_{kN_0}(v)-\d_{pN_0}\Mt_{kr}(v)\)    \nonumber\\
&+& \l\pdd{\Mt_{iN_0}(u)}{M_{pq}(u)}\bigg(
\d_{ps}\(\d_{sq}\Mt_{kN_0}(v)-\d_{qk}\Mt_{sN_0}(v)\)
-\d_{rq}\(\d_{rs}\Mt_{kN_0}(v)-\d_{pN_0}\Mt_{kr}(v) \) \bigg)        \\
&+& \e\pdd{\Mt_{kN_0}(v)}{M_{rs}(v)} \bigg(
\d_{rq}\(\d_{qs}\Mt_{iN_0}(u)-\d_{is}\Mt_{qN_0}(u) \)
-\d_{ps}\(\d_{rp}\Mt_{iN_0}(u)-\d_{rN_0}\Mt_{ip}(u) \)\bigg) \nonumber
\eqa
Thus by taking the appropriate limit with $\n\not=\m$ the left hand
side vanishes. (After noticing that $\Mt_{aN_0}(u_i,\z_i)=0$ the two
remaining terms vanish since $\pdd{\Mt_{aN_0}}{M_{Mb}}=0$.)
\bqa
\{ \Mt_{1N_0}(u_\n,\z_\n), \Mt_{1N_0}(v_\m,\e_\m) \} &=&
\{ \Mt_{1N_0}(u_\n,\z_\n), \Mt_{2N_0}(v_\m,\e_\m) \} \nonumber \\ &=&
\{ \Mt_{2N_0}(u_\n,\z_\n), \Mt_{2N_0}(v_\m,\e_\m) \} = 0 \eqa
Hence
\beq \{ u_\n,u_\m \} = \{ \z_\n,\z_\m \} = \{ u_n,\z_\m \} =0 \quad
\n\not=\m \eeq

It remains to calculate the Poisson bracket $\{ u_\n,\z_\n \}$. One
may use,
\beq \{ M_{pq}(u,\z),M_{rs}(u,\z) \} = \dd{T_{rq}}{u}T_{ps} -
\dd{T_{ps}}{u}T_{rq} \eeq
Thus
\beq
\{ \Mt_{1N_0}(u,\z),\Mt_{2N_0}(u,\z) \} = \sum_{pqrs} \(
\pdd{\Mt_{1N_0}}{M_{pq}} \pdd{\Mt_{2N_0}}{M_{rs}} - \pdd{\Mt_{2N_0}}{M_{pq}}
\pdd{\Mt_{1N_0}}{M_{rs}} \) \dd{M_{ps}}{u} ( M_{rq} + \z\d_{rq} ) \eeq
and from equation~\Ref{eq:pdc} we see,
\bqa\(
\pdd{\Mt_{1N_0}}{M_{pq}} \pdd{\Mt_{2N_0}}{M_{rs}} - \pdd{\Mt_{2N_0}}{M_{pq}}
\pdd{\Mt_{1N_0}}{M_{rs}} \)&=&
 \frac{1}{(\det M)^2} \Bigg(
-\Mt_{qp}\Mt_{sN_0}\(\Mt_{1N_0}\Mt_{2r} - \Mt_{2N_0}\Mt_{1r}\)  \nonumber\\
&& {\hskip 1.7cm} + \Mt_{sr}\Mt_{qN_0}\(\Mt_{1N_0}\Mt_{2p} -
\Mt_{2N_0}\Mt_{1p}\) \nonumber \\
&& {\hskip 2cm} + \Mt_{qN_0}\Mt_{sN_0}\(\Mt_{1p}\Mt_{2r} -
\Mt_{2p}\Mt_{1r}\) \Bigg) \label{eq:ChF}
\eqa
The last term in brackets contains three zeros and never gives a
contribution. The other terms also vanish when they are multiplied by
$\dd{M_{ps}}{u} M_{rq}$ and summed over ($ M_{rq}$ may always be
multiplied by an $\Mt$ that is not of the form $\Mt_{\#N_0}$ thereby
cancelling a $\det M$ in the denominator but leaving two zeros in the
numerator). Thus the only contribution comes from the product of the
first two terms of~\Ref{eq:ChF} and $\z\dd{M_{ps}}{u}\d_{rq}$. However
this is equal to $\z \det F_\n$.
To see this recall that,
\beq \det F_\n = \sum_{pqrs} \(
\pdd{\Mt_{1N_0}}{M_{pq}} \pdd{\Mt_{2N_0}}{M_{rs}} - \pdd{\Mt_{2N_0}}{M_{pq}}
\pdd{\Mt_{1N_0}}{M_{rs}} \) \dd{M_{pr}}{u}\d_{qs} \eeq
Thus explicit calculation shows that,
\bqa  \Bigg( \pdd{\Mt_{1N_0}}{M_{pq}} \pdd{\Mt_{2N_0}}{M_{rs}}
  - \pdd{\Mt_{2N_0}}{M_{pq}}\pdd{\Mt_{1N_0}}{M_{rs}} \Bigg)
\( \dd{M_{ps}}{u}\d_{rq} -\dd{M_{pr}}{u}\d_{qs} \)
= {\hskip 5cm} \nonumber\\
 \sum_{psr} \frac{1}{(\det M)^2}\dd{M_{ps}}{u} \Bigg(
\Mt_{1N_0}\Mt_{rN_0}\(\Mt_{sr}\Mt_{2p}-\Mt_{sp}\Mt_{2r}\)
 +\Mt_{2N_0}\Mt_{rN_0}\(\Mt_{sp}\Mt_{1r}-\Mt_{sr}\Mt_{1p}\) \nonumber\\
\hfill +\Mt_{1N_0}\Mt_{sN_0}\(\Mt_{rr}\Mt_{2p}-\Mt_{rp}\Mt_{2r}\)
+\Mt_{2N_0}\Mt_{sN_0}\(\Mt_{rp}\Mt_{1r}-\Mt_{1p}\Mt_{rr}\)  \Bigg)
 =  0 \eqa

where the terms in brackets give the required third zeros because they
are the subdeterminants of a matrix of rank 1.
Hence
\beq \{ \Mt_{1N_0}(u,\z),\Mt_{2N_0}(u,\z) \} =\z \det F_\n \eeq
thus substituting in equation \Ref{eq:pbm} we find
\beq \{ u_\n,\z_\n \} = \z_\n \eeq
as required.

\section{The Gaudin Model and other systems} \label{sec:GM}

\subsection{The Gaudin Model}
In the Gaudin model (which may be considered a degenerate case of the
magnetic chain) $T(u)$ has the following form,
\beq T(u)\equiv Z + \sum_{m=1}^M \frac{S^{(m)}}{u-\d_m} \eeq
where $S^{(m)}$ obey equation~\Ref{eq:spb} as before. $T(u)$ obeys the
following linear $R$-matrix relation,
\beq
\{ T_{\a_1\b_1}(u),T_{\a_2\b_2}(v)\} = \frac{1}{u-v} \big(
\(T_{\a_2\b_1}(u)-T_{\a_2\b_1}(v)\) \d_{\a_1\b_2} + \( T_{\a_1\b_2}(v) -
T_{\a_1\b_2}(u) \)\d_{\a_2\b_1} \big) \quad .
\eeq
Once again the spectral invariants contain the integrals of motion.
Sklyanin's conjecture for the  Gaudin model is,
\begin{conj}
 Let $\A$ and $\B$ be the same functions on $GL(N)$ as
above. Then the variables $x_j$ and $p_j$ defined by the equations,
\beq \B(T(x_i))=0, \quad\quad p_j=\A(T(x_j)) \eeq
have canonical Poisson brackets and, besides, are bound to the
Hamiltonians by the relation
\beq \det\left( p_i - T(x_i) \right) =0 \quad . \eeq
\end{conj}

This is true generically. To prove this one only needs to show that $x_i$
and $p_j$ have the correct commutation relations. The proof
works the same way as the quadratic case and will not be repeated (it
can be found in \cite{AHH:DC}).

\subsection{Other systems}

In this paper the procedure is illustrated only in the simplest cases
of the magnetic chain and the Gaudin model, however the method is much
more widely applicable.   Since the equation
$\Mt(u,\z)V_0=0$ and the calculation used to obtain $\A$ and $\B$ is
independent of the particular $R$-matrix algebra, the separation
variables thus obtained satisfy condition (1) of section~\Ref{ss:key}
(i.e. give separated equations) automatically.

 However it must be checked that they have the correct commutation
relations and at present I do not have a general proof of this.
Nevertheless it can be proved (by direct calculation) that one obtains
the correct commutation relations for systems with $R$-matrices of the
following form
\beq
R=\sum f_{ij}(u) e_{ij} \otimes e_{ji}
\eeq
provided that the $f_{ij}(u)=\frac{1}{u}+O(1)$ as $u\rightarrow 0$.
With the $e_{ij}$ as in equation~\Ref{eq:PRm} (which is clearly of this
form). This allows the result to be extended from rational to
trigonometric and elliptic $R$-matrices. Applications in these systems
and systems with dynamical $R$-matrices will be discussed in a future
publication.

\section{Conclusion}

In this paper I have shown that generically the following polynomials
in the $T$ matrix coefficients maybe used to obtain separation
variables as illustrated in the case of the magnetic chain and Gaudin model,
\bqa \A(T(u)) &=& \epsilon_{i_1 i_2 \cdots i_{N-1}} T_{i_1N}
T^2_{i_2N} \cdots  T^{N-2}_{i_{N-2}N}
\frac{T_{i_{N-1}1}}{\det(M)} \\
\B(T(u)) &=&  \epsilon_{i_1 i_2 \cdots i_{N-1}}  T_{i_1N}
T^2_{i_2N} \cdots  T^{N-2}_{i_{N-2}N}
T^{(N-1)}_{i_{N-1}N}
\eqa
in the sense that separation variables, with canonical Poisson
brackets, are given by $\B(x_i)=0$, $e^{p_i}=P_i=\A(x_i)$ for the
magnetic chain and by  $\B(x_i)=0$, $p_i=\A(x_i)$ for the Gaudin model.

With $\A$ and $\B$ in this form it is clear to see that $\B$ has an
$SL(N-1)$ symmetry corresponding to similarity transform which
leave the last row and column fixed, and that $\A$ has an $SL(N-2)$
corresponding to those transformations which leave
the first and last rows and columns fixed. Clearly in the
construction and proof it does not matter which column we choose to
call the first and last, and these associated $\A$'s and $\B$'s also give
rise to separation variables. The meaning of the symmetry and
non-uniqueness is not yet clear.

As remarked, $\B$ only gives the correct number of separation
coordinates in the generic case. For non-generic cases
the equation
\beq \tilde{M} (u,\z)V_0 =0 \eeq
must be used, it is desirable to
extend the FBA to these cases. One would need to provide a method that
told one how to decouple equation~\Ref{eq:ahh2} into an equation(s)
for the $x_i$ and an equation(s) for the $p_i$s (in terms of the $x_i$s).
The reducible systems of Kalnins et al.\cite{vbK:QR,BKM:sshl} provide
examples of such problems and the author conjectures that the notion
of irreducibility in these systems is related to this problem of
non-degeneracy.

The classical Functional Bethe Ansatz in its $2 \times 2$ version has been
successfully applied to systems with a wide variety of $R$-matrices
including dynamical $R$-matrices, as has its quantum counterpart. It is
hoped that the $SL(N)$ FBA proposed here and its quantum counterpart
might prove equally useful.

\section*{Acknowledgements}
The author would like to thank Noah Linden for invaluable support,
encouragement and discussions. And Tomasz Brzezi\'{n}ski and Alan
Macfarlane for very useful discussions.


\end{document}